\documentclass[a4paper,12pt]{article}

\usepackage{graphicx}   
\usepackage{subcaption}
\usepackage{amsmath, amssymb, physics}  
\usepackage{bm}             
\usepackage[colorlinks=true, linkcolor=blue, citecolor=blue, urlcolor=blue]{hyperref}
\usepackage[left=2cm, right=2cm, top=2cm, bottom=3cm]{geometry}
\usepackage[affil-it]{authblk}
\usepackage[center]{titlesec}
\usepackage{cite}
\usepackage{caption}

\author[1]{I.D. Ugulava}

\author[1,2]{V.I. Berezhiani}

\affil[1]{School of Physics, Free university of Tbilisi, Tbilisi 0159, Georgia and}
\affil[2]{Andronikashvili Institute of Physics (TSU), Tbilisi 0177, Georgia.}

\title{\textbf{Radiation Pressure Effect on the degenerate Electron-Positron Plasma Dynamics}}
\date{}

\begin{document}

\maketitle

\begin{abstract}
    We investigate the influence of radiation pressure on the collective dynamics of a degenerate electron–positron plasma interacting with an intense radiation flux. Assuming the Fermi energy substantially exceeds the thermal energy, we apply general covariant hydrodynamic equations to demonstrate that in a sufficiently dense plasma, radiation pressure outpaces the increase of relativistic inertia, yielding an enhanced terminal bulk Lorentz factor. Furthermore, this process is accompanied by dynamic plasma cooling.
\end{abstract}

\section{Introduction}
Radiation-driven acceleration of relativistic plasma is believed to play an important role in a variety of high-energy astrophysical environments, including active galactic nuclei, pulsar magnetospheres, gamma-ray burst sources, and compact accreting systems. In sufficiently intense radiation fields, momentum transfer between photons and plasma may efficiently accelerate charged particles and relativistic outflows. Such processes are commonly associated with the Compton rocket mechanism originally proposed by O’Dell \cite{O'Dell, Cheng} and subsequently investigated in several astrophysical contexts \cite{Sikora, Phinney}.

In the Thomson regime, the corresponding radiation pressure force may be viewed as the macroscopic manifestation of the averaged radiation reaction force acting on charged particles. In particular, the averaged Landau–Lifshitz radiation reaction force \cite{Landau_EM} reproduces the standard radiative thrust and drag terms associated with photon momentum transfer. This connection provides a unified description of radiation-driven plasma dynamics in both coherent and incoherent electromagnetic fields.

Recently, it was demonstrated that radiation reaction effects can be strongly enhanced in relativistically degenerate plasmas \cite{Berezhiani_2024}. Unlike thermally relativistic plasmas, where rapid radiative cooling suppresses relativistic particle motion \cite{Phinney}, degenerate plasmas preserve relativistic particle momenta due to the large Fermi momentum associated with extremely high densities. As a consequence, the radiation reaction force acquires a substantial density-dependent enhancement factor, which may become particularly important in compact astrophysical systems containing dense electron-positron plasma \cite{Shapiro}. For radiation intensities in the range from $10^{23}$ - $10^{30} \text{ erg} \cdot \text{s}^{-1} \cdot \text{cm}^{-2}$ relevant both for compact astrophysical sources and prospective ultra-intense laser facilities, the resulting radiative acceleration may become highly relativistic.

While the general covariant hydrodynamic formulation of radiation reaction in degenerate plasma has recently been established \cite{Berezhiani_2024}, the resulting collective plasma dynamics in realistic radiation configurations remains largely unexplored. In particular, it remains unclear how the degeneracy-enhanced radiation force modifies the bulk acceleration of an electron-positron fluid, whether the terminal flow velocity differs from the corresponding single-particle dynamics, and under what conditions collective effects become dominant.

In the present work, we investigate the dynamics of a relativistically degenerate electron--positron plasma interacting with an axially symmetric radiation field. Starting from the relativistic fluid equations including the radiation reaction force, we derive the evolution equations governing the bulk plasma motion and obtain analytical expressions for the terminal velocity and asymptotic Lorentz factor. We demonstrate that, whereas standard radiation-reaction models typically converge to a single-particle terminal velocity, the collective dynamics of a degenerate plasma subject to varying radiation intensity lead to qualitatively different asymptotic states. In sufficiently dense plasmas, the resulting Lorentz factor may substantially exceed the corresponding single-particle Landau--Lifshitz limit due to the degeneracy-enhanced radiation reaction force.

The physical conditions considered in this work may arise transiently in compact astrophysical environments characterized by intense pair production, including pulsar magnetospheres and magnetar flares~\cite{Sturrock, Beloborodov}. Intense electron--positron pair creation also occurs during the gravitational collapse of massive stars~\cite{Ruffini_2012}. Furthermore, superdense electron--positron (e-p) plasmas may exist in GRB sources, where the plasma density can reach the range $n=(10^{30}\text{--}10^{34})\text{~cm}^{-3}$~\cite{Ruffini_2010}.

Similar regimes may also become accessible in next-generation ultra-intense laser facilities capable of producing dense e-p pair plasmas~\cite{Berezhiani_2015}. Dense e-p plasmas may soon be produced under laboratory conditions as well. Modern petawatt laser systems are already capable of producing ultrashort pulses with focal intensities of $I = 2 \cdot 10^{22}\text{~W/cm}^2$~\cite{Yanovsky}. Pulses with even higher intensities, exceeding $I = 10^{26}\text{~W/cm}^2$, are likely to become available in the near future, either in the laboratory or in Lorentz-boosted frames~\cite{Dunne}. The interaction of such pulses with gaseous or solid targets could lead to the generation of e-p plasmas with above-solid-state densities in the range of $(10^{23}\text{--}10^{28})\text{~cm}^{-3}$~\cite{Wang}.

At sufficiently high densities, the thermodynamic properties of the plasma may differ qualitatively from those of a classical, non-degenerate plasma. In particular, when the average interparticle distance becomes smaller than the thermal de Broglie wavelength, the plasma behaves as a degenerate Fermi gas. In this regime, the wave nature of the particles becomes important and the quantum statistical effects associated with the Pauli exclusion principle must be taken into account. At the same time, mutual interactions between the plasma particles become less important as the density increases, allowing the plasma to be treated as increasingly ideal. If the thermal energy of the particles (electrons and positrons) is much lower than their Fermi energy, the plasma may be treated as cold and degenerate.

\section{Relativistic Fluid Formalism for Degenerate Plasma}
In this section, we briefly summarize the relativistic fluid formalism for a degenerate electron–positron plasma in the presence of radiation reaction effects. The general covariant hydrodynamic model was previously derived in Ref.\cite{Berezhiani_2024} for arbitrary temperature and degeneracy. Here, we present the formulation in a form convenient for the subsequent analysis of one-dimensional plasma dynamics and explicitly express the thermodynamic quantities through generalized Fermi–Dirac integrals. We consider an electron–positron plasma subjected to an intense radiation field in the Thomson regime, where the photon energy in the particle rest frame remains much smaller than the electron rest-mass energy. Under this condition, the radiation force may be described classically using the Landau–Lifshitz radiation reaction force. Assuming local thermodynamic equilibrium, the plasma is described by the Fermi–Dirac distribution function:
\begin{equation}
    f = \frac{2}{(2\pi\hbar)^3} \left[ \exp \left( \frac{ p_\alpha U^\alpha - \mu}{T} \right) + 1 \right]^{-1},
\end{equation}
where $p^\alpha = mcu^\alpha$ denotes the particle four-momentum and $u^\alpha = [\gamma, \gamma\mathbf{v}/c]$ is the particle reduced four-velocity, with $\gamma = (1 - \mathbf{v}^2/c^2)^{-1/2}$ being the Lorentz factor. Given the metric signature $g_{\alpha \beta} = \text{diag}(1, -1, -1, -1)$, the four-velocity satisfies the normalization $u^\alpha u_\alpha = 1$. The plasma bulk four-velocity is defined as $U^\alpha = [\Gamma c, \Gamma \mathbf{V}]$, where $\mathbf{V}$ is the macroscopic flow velocity and $\Gamma = (1 - \mathbf{V}^2/c^2)^{-1/2}$ is the bulk Lorentz factor. The parameters $\mu$ and $T$ represent the chemical potential and temperature of the system, respectively. The dynamical evolution of the distribution function is governed by the relativistic collisionless Boltzmann equation, modified to account for radiative effects:
\begin{equation}
    p^{\alpha} \frac{\partial f}{\partial x^{\alpha}} + \frac{\partial}{\partial p^{\alpha}} \left(\left[\frac{e}{c}F^{\alpha\beta}p_\beta +mc g^{\alpha} \right]f \right) = 0
\end{equation}
Where $F^{\alpha \beta}$ is the electromagnetic field tensor  and $g^\alpha$ is the radiation reaction four-force, which was originally derived by Landau and Lifshitz \cite{Landau_EM} (see also \cite{Berezhiani_2004, Berezhiani_2008}):
\begin{equation}
    g^{\alpha} = \frac{2e^3}{3m c^3} \partial_\gamma F^{\alpha\beta} u^\gamma u_\beta + \frac{\sigma}{c} \left( \bar{T}^{\alpha\gamma} u_\gamma - \bar{T}^{\gamma\lambda} u_\gamma u_\lambda u^{\alpha} \right),
\end{equation}
where $\sigma = \frac{8\pi e^4}{3m^2c^4}$ denotes the Thomson cross-section. The term $\bar{T}^{\mu\nu}$ represents the energy-momentum tensor of the electromagnetic field, given by:
\begin{equation}
    \bar{T}^{\mu\nu} = \frac{1}{4\pi} \left( F^{\mu\alpha} F^\nu_{\ \alpha} - \frac{1}{4} g^{\mu\nu} F_{\alpha\beta} F^{\alpha\beta} \right).
\end{equation}
The time average of the electro-magnetic force and the first term in the expression for $g^\alpha$ vanish for rapidly oscillating fields. Consequently, the leading-order contributions to the radiation pressure are governed by the second-order terms of the electromagnetic field within the energy-momentum tensor. Now, we define the \textbf{particle four-flux} and the \textbf{energy-momentum tensor} as the first and second moments of the distribution function $f$, respectively:
\begin{equation}
    N^{\alpha} = c \int \frac{d^3 \mathbf{p}}{p^0} p^{\alpha} f
\end{equation}
\begin{equation}
    T^{\alpha\beta} = c \int \frac{d^3 \mathbf{p}}{p^0} p^{\alpha} p^{\beta} f
\end{equation}
By taking the moments of the relativistic Boltzmann equation \cite{Berezhiani_2024}, the averaged dynamical equation for the plasma can be derived as:
\begin{equation}
    \frac{\partial T^{\mu\nu}}{\partial x^{\nu}} = m c^2 \langle g^{\mu} \rangle = F_{\text{rad}}^\mu
\end{equation}
where the averaged force density is given by:
\begin{equation}
    \langle g^\alpha \rangle = \int \frac{d^3 \mathbf{p}}{p^0} g^\alpha f
\end{equation}
The resulting radiation four-force $F_{\text{rad}}^\mu$, representing the momentum transfer between the field and the plasma, is expressed as:
\begin{equation}
    F_{\text{rad}}^\mu = \frac{\sigma}{c} \bar{T}^{\mu \nu} N_\nu - \frac{\sigma}{m^2 c^3} \bar{T}_{\nu\lambda} M^{\nu\lambda \mu}
\end{equation}
where $M^{\nu\lambda\mu}$ denotes the third moment of the distribution:
\begin{equation}
    M^{\nu\lambda\mu} = c \int \frac{d^3 \mathbf{p}}{p^0} p^{\nu} p^{\lambda} p^{\mu} f
\end{equation}
Since the moments of the distribution are completely dependent on the metric on the space and the four-flow, we can linearly decompose the moments using the following methods:
\begin{equation}
    N^\alpha = nU^\alpha, \quad T^{\alpha \beta} = (\varepsilon + P)\frac{U^\alpha U^\beta}{c^2} - g^{\alpha\beta} P, \quad M^{\mu\nu\lambda} = A_1U^\mu U^\nu U^\lambda + A_2 g^{\{\mu \nu}U^{\lambda\}}
\end{equation}
where $n$, $\varepsilon$, and $P$ are the particle density, internal energy density, and pressure, respectively, in the rest frame of the plasma. While these parameters were demonstrated for a plasma of any degree of degeneracy in Ref. \cite{Berezhiani_2024}, we express them here in terms of the more standard Fermi-Dirac integrals:
\begin{equation}
    n = \frac{\sqrt{2}m^3c^3 \theta^{\frac{3}{2}}}{\pi^2\hbar^3}[I_{1/2}(\eta, \theta) + \theta I_{3/2}(\eta, \theta)]
\end{equation}
\begin{equation}
    P =\frac{2\sqrt{2}m^4c^5\theta^{5/2}}{3\pi^2\hbar^3}\left[I_{3/2}(\eta,\theta) + \frac{\theta}{2}I_{5/2}(\eta,\theta)\right]
\end{equation}
\begin{equation} 
    \varepsilon = nmc^2 + \frac{\sqrt{2}\, m^{4} c^{5} \theta^{5/2}}{\pi^{2}\hbar^{3}} \left[ I_{3/2}(\eta,\theta) + \theta I_{5/2}(\eta,\theta) \right]
\end{equation}
where $\theta = T/mc^2$ is the reduced temperature, $\eta = (\mu-mc^2)/T$ is the reduced chemical potential, and $I_k(\eta, \theta)$ is the generalized Fermi-Dirac integral \cite{Landau_ST}:
\begin{equation}
    I_k(\eta,\theta) = \int_0^\infty\frac{x^k \sqrt{1 + \frac{\theta x}{2}}}{\exp(x - \eta) + 1}dx
\end{equation}
As shown by Faussurier \cite{Faussurier}, the same results can be obtained by considering the grand potential of the Fermi-Dirac gas and calculating the thermodynamic properties of the plasma. Following the Ref.\cite{Faussurier}, one can determine the entropy per particle using thermodynamic relation between the grand potential and the entropy:
\begin{equation}
    S = - \eta + \frac{\frac{5}{3} I_{3/2}(\eta, \theta) + \frac{4\theta}{3} I_{5/2}(\eta, \theta)}{I_{1/2}(\eta, \theta) + \theta I_{3/2}(\eta, \theta)}
\end{equation}
The coefficients for the third moment are found to be:
\begin{equation}
    A_1 = \frac{\sqrt{2} m^5c^3 \theta^{\frac{3}{2}}}{\pi^2 \hbar^3} [I_{1/2}(\eta, \theta) + 5\theta I_{3/2}(\eta, \theta) + 6 \theta^2 I_{5/2}(\eta, \theta) + 2\theta^3 I_{7/2}(\eta, \theta)]
\end{equation}
\begin{equation}
    A_2 = -\frac{\sqrt{2}m^5c^5 \theta^{\frac{3}{2}}}{3\pi^2 \hbar^3}[2\theta I_{3/2}(\eta, \theta) + 3 \theta ^2 I_{5/2}(\eta, \theta) + \theta ^3I_{7/2}(\eta, \theta)]
\end{equation}
which allows the radiation-force term to be expressed in the form
\begin{equation}
    F_{rad}^{\mu} = \frac{\sigma}{c}\bar{T}^{\mu\nu}U_{\nu}\left(n - \frac{1}{m^2c^2}2A_2\right) - \frac{\sigma}{m^2c^3}A_1\bar{T}^{\nu\lambda}U_{\nu}U_{\lambda}U^{\mu}
\end{equation}
Using Eq. (19) together with Eq. (7), the fluid-dynamical equation can be written in contravariant form as
\begin{equation}
     w\frac{U^\nu}{c^2} \partial_\nu U^\mu  - \left(g^{\mu\nu} - \frac{U^\mu U^\nu}{c^2}\right)\partial_\nu P = R^\mu
\end{equation}
where $w=\varepsilon+P$ denotes the enthalpy density and $R^\mu$ represents the contribution arising from the radiation force,
\begin{equation}
    R^\mu = \frac{\sigma}{c}\left(n - \frac{1}{m^2c^2}2A_2\right) \left(\bar{T}^{\mu\nu}U_{\nu} - \frac{1}{c^2}\bar{T}^{\nu\lambda}U_{\nu}U_{\lambda} U^{\mu}\right)
\end{equation}
Additional insight is obtained by projecting Eq. (7) onto the fluid four-velocity. Multiplication by $U_\mu$ yields the scalar relation
\begin{equation}
    \partial_\nu \left(w U^\nu\right)   -U^{\nu}\partial_\nu P = \frac{\sigma}{c}\bar{T}^{\mu\nu}U_{\nu}U_\mu \left(n - \frac{1}{m^2c^2}2A_2 -\frac{1}{m^2}A_1\right) 
\end{equation}
Employing the thermodynamic identity
\begin{equation}
    d\left(\frac{w}{n}\right) -\frac{1}{n}dP  = TdS 
\end{equation}
together with the continuity equation 
\begin{equation}
    \partial_\mu N^\mu = \partial_\mu(nU^\mu) = 0
\end{equation}
the left-hand side of (21) can be recast as
\begin{equation}
     \partial_\nu \left(w U^\nu\right)   -U^{\nu}\partial_\nu P= n\Gamma\left(\frac{d}{dt}\left(\frac{w}{n} \right) - \frac{1}{n}\frac{dP}{dt}\right) = n\Gamma T\frac{dS}{dt}
\end{equation}
Consequently, the entropy evolution equation takes the form
\begin{equation}
    nmc^2\Gamma\theta \frac{dS}{dt} = \frac{\sigma}{c}\bar{T}^{\mu\nu}U_{\nu}U_\mu \left(n - \frac{1}{m^2c^2}2A_2 -\frac{1}{m^2}A_1\right) 
\end{equation}
Making use of the relation
\begin{equation}
    \frac{1}{n\theta}  \left(n - \frac{1}{m^2c^2}2A_2 -\frac{1}{m^2}A_1\right)  = -\frac{4}{3\theta}\frac{2\theta I_{3/2} + 3 \theta^2 I_{5/2} + \theta^3I_{7/2}}{I_{1/2} + \theta I_{3/2}}
\end{equation}
the entropy change rate can be written as
\begin{equation}
    \frac{dS}{dt} = -\frac{4}{3\theta}\frac{\sigma}{mc^3\Gamma}\bar{T}^{\mu\nu}U_{\nu}U_\mu \left[\frac{2\theta I_{3/2} + 3 \theta^2 I_{5/2} + \theta^3I_{7/2}}{I_{1/2} + \theta I_{3/2}}\right]
\end{equation}
This expression shows that the entropy per particle decreases during radiation-driven acceleration. Since the entropy is a monotonic function of temperature for the present equation of state, the result implies a reduction of the plasma temperature as the bulk flow gains kinetic energy. The radiation field therefore acts not only as a source of momentum but also as an effective cooling mechanism for the accelerating plasma.  

We now consider the limit of a completely degenerate electron-positron plasma, corresponding to the zero-temperature limit $T \rightarrow 0$ (or $\theta \rightarrow 0$). In this regime, the Fermi--Dirac distribution reduces to a step function, allowing all distribution moment integrals to be evaluated analytically. Under these conditions, the chemical potential coincides with the Fermi energy. We define the dimensionless Fermi momentum $\chi$ as:
\begin{equation}
\chi = \frac{p_F}{mc} = \left(\frac{n}{n_c}\right)^{\frac{1}{3}} .
\end{equation}
where $n_c = \frac{m^3c^3}{3\pi^2\hbar^3} = 5.9 \times 10^{29} cm^{-3}$. One can also show the relation $\chi=\sqrt{\left(\frac{\mu}{mc^2}\right)^2 - 1}$. Now, following the reference, in this limit, Eq. (7) reduces to the following contravariant form:
\begin{equation}
nm\gamma_F U^\beta \frac{\partial U^\alpha}{\partial x^\beta} 
- \left( g^{\alpha \beta} - \frac{1}{c^2} U^\alpha U^\beta \right) \frac{\partial P}{\partial x^\beta} 
= R^\alpha,
\end{equation} 
where $\gamma_F = \sqrt{1+\chi^2}$ and $R^{\mu}$ reduces to the following form:
\begin{equation}
R^{\mu}
= \frac{\sigma n}{c}\left(1+\frac{2}{5}\chi^2\right)
\left[\bar{T}^{\mu\nu}U_\nu
-\frac{1}{c^2}\bar{T}^{\nu\lambda}U_\nu U_\lambda U^\mu \right].
\end{equation}
As demonstrated in the \cite{Berezhiani_2024, Faussurier}, in the fully degenerate limit, one should express pressure
\begin{equation}
    P =\frac{m^4c^5}{24\pi^2\hbar^3} \left[3\ln{\left(\sqrt{\chi^2 + 1} + 1\right)} + \sqrt{1 + \chi^2} \left(2\chi^3 -3\chi \right)\right] + \mathcal{O}\left( \theta ^2\right)
\end{equation}
The general expression for the entropy per particle $S$, as defined via generalized Fermi--Dirac integrals in Eq.~(16), identically vanishes in the strict zero-temperature limit ($T \rightarrow 0$ or $\theta \rightarrow 0$), conforming to the third law of thermodynamics. In formulating the hydrodynamic evolution of the plasma, the fluid parameters are expanded about the zero-temperature degenerate baseline. This perturbative framework remains physically consistent provided the plasma temperature remains well below the characteristic Fermi temperature ($\theta \ll \theta_F$). 

\begin{figure}
    \centering
    \includegraphics[width=0.5\linewidth]{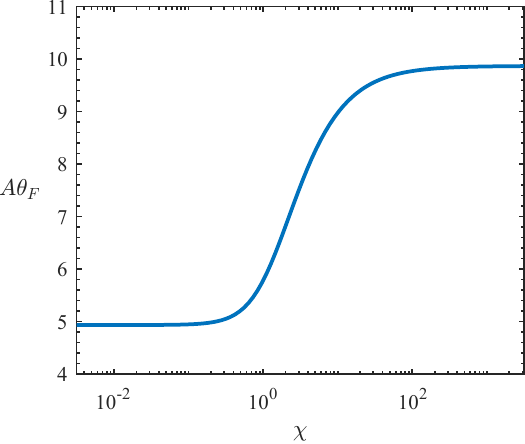}
    \caption{Dependence of $A(\chi)\theta_F(\chi)$ on the $\chi$ }
    \label{fig:A_theta_F}
\end{figure}

Crucially, as the plasma parcel undergoes radiation-driven acceleration, its macroscopic thermodynamic state evolves. To justify the omission of higher-order thermal corrections in the equations of motion---specifically regarding the rest-frame pressure and particle density---the temperature dependence of the entropy must be thoroughly evaluated. As established in \cite{Faussurier}, the low-temperature thermodynamic expansion of the entropy per particle can be structurally constrained to linear term of the reduced temperature:
\begin{equation}
    S = A(\chi)\theta
\end{equation}
where $A(\chi)$ is an expansion coefficient, the lengthy explicit expression of which was previously derived in Ref.~\cite{Faussurier}. Rather than repeating these algebraic details here, we illustrate the functional dependence of $A(\chi)\theta_F(\chi)$ on the Fermi momentum $\chi$ graphically in Figure~\ref{fig:A_theta_F} This specific ratio is chosen because $A(\chi) \to \infty$ while $\theta_F(\chi) \to 0$ as $\chi \to 0$, providing a well-behaved, normalized quantity.

Substituting this approximation into Eq.~(28) yields a remarkably simple governing equation for the temperature rate of change:
\begin{equation}
    \dot\theta = -\frac{4\chi^2}{5\theta A\left(\chi\right)}\frac{\sigma}{mc^3\Gamma}\bar{T}^{\mu\nu}U_{\nu}U_\mu 
\end{equation}

\section{1D Dynamics in an Axially Symmetric Radiation Field}
Having demonstrated the Compton rocket effect for a degenerate electron-positron fluid, we now evaluate the dynamics for specific, physically relevant radiation profiles. For simplicity, we assume that the radiation-reaction force (RRF) is the sole driver of the electron--positron plasma, with the particle number density and pressure in the rest frame remaining constant during the acceleration process. Under this approximation, the equation of the motion simplifies to the following expression:
\begin{equation}
   nm\gamma_F U^\beta \frac{\partial U^\alpha}{\partial x^\beta} = R^\alpha.
\end{equation}

Consider a plasma parcel propagating along the symmetry axis of an axially symmetric radiation field. In this configuration, the radiation energy--momentum tensor is characterized by the following non-vanishing components: $\bar{T}_{00} = U$ (energy density), $\bar{T}_{03} = \bar{T}_{30} = F$ (energy flux), and the radiation pressure components $\bar{T}_{11} = \bar{T}_{22} = P_{\perp}$ (transverse) and $\bar{T}_{33} = P_{\parallel}$ (parallel). By evaluating the components of the radiation-reaction four-force $R^\alpha$ within this geometry, we obtain:
\begin{equation}
R^3 = \sigma n\left(1 + \frac{2}{5}\chi^2 \right) \Gamma^3\left(F\beta^2 -(P_{\parallel} + U)\beta + F\right)
\end{equation}
where $\beta = V/c$ is the reduced speed of the flow. From this, we obtain the 1D fluid dynamics equation:
\begin{equation}
    \dot \beta =  \frac{\sigma }{m\gamma_Fc \Gamma} \left(1 + \frac{2}{5}\chi^2 \right) \left(F\beta^2 -(P_{\parallel} + U)\beta +F\right)
\end{equation}
with the corresponding terminal velocity:
\begin{equation}
\beta_{cr} = \frac{U + P_{\parallel}}{2F}\left(1 \pm \sqrt{1-\frac{4F^2}{(U+P_{\parallel})^2}}\right) = \lambda \pm \sqrt{\lambda^2 - 1}
\end{equation}
where
\begin{equation}
    \lambda = \frac{U + P_{\parallel}}{2F} \ge 1
\end{equation}
Notably, the restriction imposed on the parameter $\lambda$ guarantees a unique critical velocity, defined by $\beta_{cr} = \lambda - \sqrt{\lambda^2 -1}$. The last result demonstrates excellent consistency with the single-particle dynamics previously established by Sikora and Kovner \cite{Sikora, Kovner}. Furthermore, the degenerate gas is observed to asymptotically approach this identical terminal velocity. This equivalence aligns with physical intuition: because each constituent particle is subjected to uniform force-balance conditions, the ensemble average necessarily converges to the same terminal state. Consequently, the macroscopic flow velocity is guaranteed to align with the single-particle trajectory. Following Eq. (34)
\begin{equation}
    \dot\theta = -\frac{4\chi^2}{5\theta A\left(\chi\right)}\frac{\sigma \Gamma}{mc} \left(\beta^2P_{\parallel} + 2\beta F + U\right) 
\end{equation}
one can relate the temperature and the velocity with the help of Eq.(37):
\begin{equation}
    \kappa \theta\frac{d\theta}{d\beta} = - \frac{1}{1 - \beta^2}\frac{\left(\beta^2P_{\parallel} + 2\beta F + U\right)}{\left(F\beta^2 -(P_{\parallel} + U)\beta + F\right)} 
\end{equation}
where 
\begin{equation}
    \kappa = \frac{5A\left(\chi\right)\left(1 + \frac{2}{5}\chi^2 \right) }{4\chi^2\sqrt{1+\chi^2}} 
\end{equation}

Consequently, integration of both sides of the expression yields the explicit velocity-temperature relation:
\begin{equation}
    \kappa \theta^2 + \ln{\left(\frac{1-\beta^2}{1-2\lambda\beta + \beta^2}\right)} +\lambda'\ln{\left(\frac{\lambda +\sqrt{\lambda^2 - 1} -\beta}{\lambda -\sqrt{\lambda^2 -1} - \beta} \right)} = \kappa \theta_0^2 + \lambda' \ln{\left(\frac{\lambda + \sqrt{\lambda^2 -1}}{\lambda - \sqrt{\lambda^2 -1}} \right)}
\end{equation}
where $\lambda' =\frac{U - P_{\parallel}}{2F}$ and ${\theta_0}$ is the initial temperature, when the parcel has the zero velocity. 
\begin{figure}[!h]
    \centering
    \includegraphics[width=0.5\linewidth]{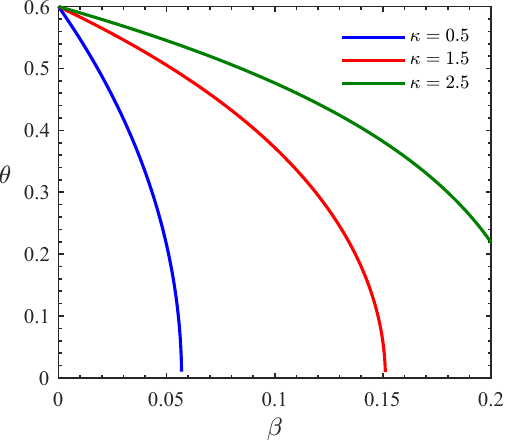}
    \caption{Temperature dependence on the flow velocity for the different $\kappa$ values. $\lambda=2$, $\lambda' = 0.5$}
    \label{fig:theta_vs_beta}
\end{figure}
It is evident that higher-density plasmas undergo significantly more rapid thermal quenching. 

Conversely, since the thermal evolution is decoupled from the macroscopic dynamics of the plasma, the hydrodynamic equations of motion can be evaluated independently. This decoupling permits analytical solutions in several asymptotic limits where the spatial configuration of the radiation field simplifies. A primary example is the ideal plane-wave limit, where the radiation field components satisfy $U = F = P_{\parallel}$ alongside vanishing transverse pressure ($P_{\perp} = 0$). Under these geometric constraints, Eq.~(37) reduces to 
\begin{equation}
    \bar{T}^{\mu\nu} = F n^\mu n^\nu,
\end{equation}
where $n^\mu = (1, 0, 0, 1)$ is the null vector indicating the direction of radiation ($n^\mu n_\mu = 0$), and $F = W/c$ represents the radiation energy flux, with $W$ being the radiation intensity. Assuming the electron--positron plasma flows exclusively along the direction of radiation propagation we obtain Landau-like dynamic system for the degenerate plasma
\begin{equation}
   \dot{\beta} = \frac{\sigma F \left(1+\frac{2}{5}\chi^2\right)}{m \gamma_F c \Gamma^3} \frac{1-\beta}{1+\beta} 
\end{equation}
It is evident from this expression that $\dot{\beta} > 0$ for all $\beta < 1$, implying that the plasma undergoes continuous acceleration in the direction of the radiation flux. This result recovers the classical Landau-Lifshitz solution in the limit where $\chi \rightarrow 0$, where the equation of motion can be solved directly. However, to better represent physical astrophysical environments, we shall apply a significantly more realistic model in the subsequent chapter. In realistic astrophysical scenarios, the radiation energy density $F$ is not constant but varies with the distance from the source. The simplest model for this dependence is the inverse-square law:
\begin{equation}
    F(z) = \frac{W_0 R^2}{c(R+z)^2}
\end{equation}
where $z$ is the distance from the source, $R$ is the characteristic radius of the source, and $W_0$ is the radiation intensity at $z=0$. To analyze the evolution of the plasma, it is more convenient to reformulate the equation of motion in terms of the Lorentz factor $\Gamma$ and the spatial coordinate $z$, rather than time. Using the relation $\dot{\beta} = c\beta \frac{d\beta}{dz}$, the equation of motion transforms to:
\begin{equation}
   \frac{1+\beta}{1-\beta}\gamma^3 \beta\frac{d\beta}{dz} = \frac{\sigma W_0 \left(1+\frac{2}{5}\chi^2\right)}{m \gamma_F c^3} \frac{R^2}{(R+z)^2}
\end{equation}
Substituting these relations into the equation of motion, we obtain a simplified differential equation for the Lorentz factor:
\begin{equation}
      \left(\Gamma + \sqrt{\Gamma^2-1}\right)^2\frac{d\Gamma}{dz} = K \frac{R}{(R+z)^2}
\end{equation}
where $K$ is a dimensionless parameter characterizing the coupling between the radiation field and the degenerate plasma:
\begin{equation}
    K = \frac{\sigma W_0 \left(1+\frac{2}{5}\chi^2\right)}{m \gamma_F c^3} R.
\end{equation}

Integrating both sides from the source ($z=0, \gamma_v=1$) to a distance $z$, we find:
\begin{equation}
    \frac{2}{3}(\Gamma^2-1)^{3/2} + \frac{2}{3}\Gamma^3 -\Gamma + \frac{1}{3} = K\frac{z}{R+z}
\end{equation}
The Lorentz factor $\Gamma_f$ gained by the plasma as it moves to infinity ($z \to \infty$) is determined by:
\begin{equation}
    \frac{2}{3}(\Gamma_f^2-1)^{3/2} + \frac{2}{3}\Gamma_f^3 -\Gamma_f + \frac{1}{3} = K
\end{equation}
The analytical solution for $\gamma_v$ can be expressed as:
\begin{equation}
    \Gamma_f = \frac{1}{2} \left[ D_1 + D_2 + \frac{1}{D_1+D_2} \right],
\end{equation}
where:
\begin{equation}
    D_1 = \sqrt[3]{3K - 1 + \sqrt{9K^2 - 6K}}, \quad D_2 = \sqrt[3]{3K - 1 - \sqrt{9K^2 - 6K}}.
\end{equation}
In the limit of strong coupling ($K \gg 1$), the terminal Lorentz factor follows the asymptotic scaling:
\begin{equation}
    \Gamma_f \approx \left(\frac{3}{4}K\right)^{1/3}.
\end{equation}

\begin{figure}[!h]
    \centering
    \includegraphics[width=0.5\linewidth]{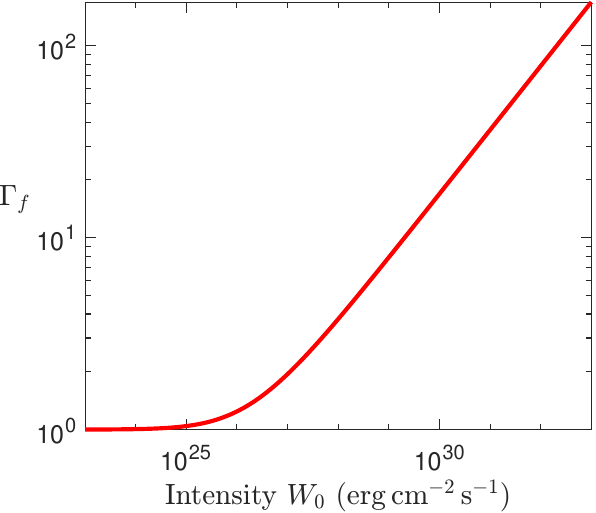}
    \caption{Dependence of the $\Gamma_f$ on th intensity $W_0$ for the fixed density $n = 10^{32} \, \mathrm{cm^{-3}}$.}
    \label{fig:intensity_vs_gamma}
\end{figure}

\begin{figure}[!h]
    \centering
    \includegraphics[width=0.5\linewidth]{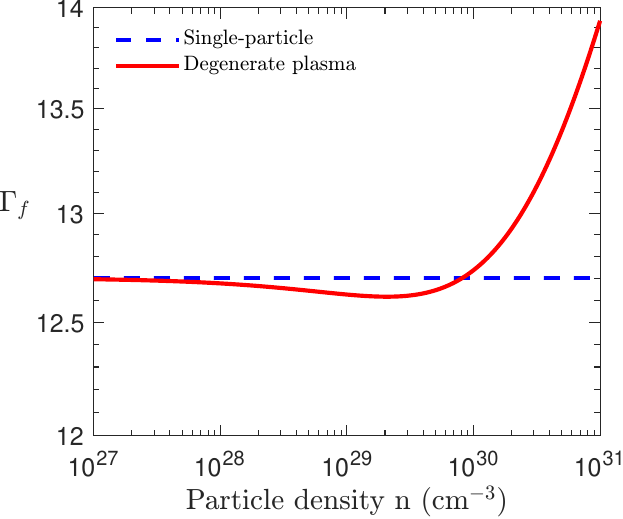}
    \caption{Dependence of $\Gamma_f$ on the plasma density for the fixed radiation intensity $W_0 = 10^{30} \, \mathrm{erg \, cm^{-2} \, s^{-1}}$. The dashed blue line depicts classical Landau-Lifshitz limit for the single particle}
    \label{fig:density_vs_gamma}
\end{figure}

As illustrated in Figure~\ref{fig:intensity_vs_gamma}, an increase in radiation intensity leads to a dramatic enhancement of the plasma energy. However, more complex phenomenon emerges when examining the role of plasma density (Figure~\ref{fig:density_vs_gamma}). By modulating the concentration of the plasma, we can directly control the degeneracy parameter $\chi$. Crucially, in regimes where the condition $1 + \frac{2}{5}\chi^2 < \gamma_F$ holds, the increase in the effective mass of the particles within the degenerate fluid outweighs the additional momentum imparted by the enhanced radiation-reaction force. Consequently, in this high-density limit, the total Lorentz factor acquired by the collective plasma becomes smaller than that of a single particle subjected to the same radiation field. Once the plasma concentration is increased beyond the threshold of $\chi > \frac{\sqrt{5}}{2}$, the Lorentz factor is expected to reach values approximately $1.5$ to $3$ times higher than those observed in the non-degenerate regime. In this high-density limit, the enhancement of the radiation-reaction force significantly outpaces the increase in relativistic inertia. This transition marks the regime where quantum effects become the dominant mechanism for energy transfer, potentially explaining the highly-relativistic outflows observed in compact astrophysical sources where traditional acceleration models fall short.

\section{Conclusion}

In this work, we have established a fully relativistic fluid model incorporating radiation-reaction effects to investigate the dynamics and thermodynamics of a degenerate electron--positron plasma. A key finding of our thermodynamic analysis is that radiation-driven acceleration is intrinsically coupled with a rapid thermal quenching of the plasma parcel. As the macroscopic flow absorbs energy from the radiation field and undergoes collective relativistic acceleration, thermal energy is continuously extracted from the rest frame of the plasma, leading to a monotonic decrease in entropy per particle. Because entropy scales linearly with temperature in this low-temperature regime ($\theta \ll \theta_F$), this radiative momentum transfer acts as a highly efficient cooling mechanism. Consequently, the plasma parcel rapidly cools as it accelerates toward its terminal velocity. This thermal quenching suppresses pressure-driven dispersion, ensuring that the accelerating jet remains highly collimated, cold, and dynamically stable even in high-intensity radiation environments.

Furthermore, for an axially symmetric radiation field, we derived the explicit velocity-temperature relation, demonstrating that the fluid asymptotically approaches a unique, stable terminal velocity $\beta_{\text{cr}} = \lambda - \sqrt{\lambda^2 - 1}$. This state is structurally identical to the single-particle trajectory established by Sikora and Kovner, confirming that under uniform force-balance conditions, the ensemble average converges to the single-particle equilibrium state.

In a strong radiation field, a dense fluid of particles moves collectively as a group, completely differently than a single, isolated particle would. We demonstrate the precise mathematical solution for how dense plasma accelerates when blasted by intense radiation. In the high-density regime exceeding the critical Fermi momentum threshold $\chi > \frac{\sqrt{5}}{2}$, the degeneracy-enhanced radiation-reaction force outpaces the increased fluid inertia, yielding terminal Lorentz factors that are a factor of several times higher than those predicted by classical, non-degenerate, or single-particle models. Conversely, below this threshold, the increased effective mass dominates, suppressing the macroscopic acceleration relative to the single-particle baseline.

These results demonstrate that the macroscopic evolution of dense pair plasmas cannot be treated merely as a collection of isolated single-particle trajectories. The amplification of the radiative coupling provides a compelling mechanism for generating highly degenerated outflows, which could be relevant to compact, high-energy astrophysical environments.

\section*{Acknowledgment}
The research was supported by the Shota Rustaveli National Science Foundation grant No. FR-24-1751. The
research of I.D.U. was supported by the Knowledge Foundation at the Free University of Tbilisi.

\bibliographystyle{unsrt}

\end{document}